\documentclass[12pt]{article}
\usepackage{sectsty}
\sectionfont{\fontsize{12}{15}\selectfont}
\usepackage[top = 2.5cm, bottom = 2.0cm, left = 2cm, right = 2cm]{geometry}
\usepackage[T1]{fontenc}
\usepackage[utf8]{inputenc}
\usepackage{amsmath}
\usepackage{mathtools}
\usepackage[toc,page]{appendix}
\usepackage[greek,english]{babel}
\usepackage{authblk}
\usepackage{physics}
\usepackage{ragged2e}

\begin{document}
\title{Quantum Entanglement and Axion Physics}
\author{A. Nicolaidis}
\affil{Theoretical Physics Department \\
Aristotle University of Thessaloniki, 54124 Thessaloniki, Greece\\
}
\date{}

\maketitle

e-mail address: nicolaid@auth.gr
\begin{abstract}

The axion particle is the outcome of the proposed Peccei-Quinn mechanism for solving the strong CP problem. Axion is also a popular dark matter candidate. Thus there is an increased interest in establishing its existence. Axions couple to two photons and most experiments search for the transition of an axion into a photon, in the presence of a magnetic field. In our study we examine the coupling of the axion into a pair of entangled photons. The presence of a magnetic field changes the polarization correlations of the entagled photons, thus offering an unambiguous signature for axion existence. 
\end{abstract}

The Standard Model of particle physics offers a successful description of the fundamental constituents of matter and the interactions among them. The discovery of the Higgs boson \cite{1} confirmed our ideas about the mechanism of spontaneous symmetry breaking and the generation of masses of the weak gauge bosons. Yet there are unresolved issues and one of them is the strong CP problem. The strong CP problem arises from the non-Abelian nature of QCD. QCD vacuum allows the existence of a CP-violating term 

\begin{equation}\label{eq1}
\begin{aligned}
    L_{\theta} = \theta \: \frac{g_3}{32\pi^2}\:G^{\mu\nu}_{\alpha}\:\tilde{G}_{\alpha\mu\nu}
\end{aligned}
\end{equation}
where $g_3$ is the coupling constant of the strong force, $G_{\mu\nu}$ is the gluon field strength tensor, $\tilde{G}_{\mu\nu}$ is its dual and $\theta$ is the angle determining the strength of CP violation. From the absence of a measurable neutron electric dipole moment we infer that $\theta < 10^{-9}$. Here lies the strong CP problem: why this angle, is exceedingly small?

One may accept the small $\theta$ value by fiat, but Peccei and Quinn offered an elegant solution based on spontaneous symmetry breaking \cite{2}. A global chiral symmetry, known as U(1)$_{PQ}$, is introduced. This symmetry is spontaneously broken when a scalar field develops a vacuum expectation value. The associated Goldstone boson couples to two gluons and at the minimum of the effective potential the total coefficient of $G\tilde{G}$ (shorthand of $G^{\mu\nu}_{\alpha}\;\tilde{G}_{\alpha\mu\nu}$) relaxes to zero. The physical particle, called the axion \cite{3,4}, replaces effectively $\theta$ and the Lagrangian no longer has a CP-violating term. The axion is also one of the leading candidates for dark matter \cite{5,6}. It appears then that the search for axions is a highly important issue for fundamental physics.

Similar to the axion-two gluon coupling there is an axion-two photon coupling 

\begin{equation}\label{eq2}
    \begin{aligned}
        L_{a\gamma\gamma} = -\frac{1}{4}\: g \: a \: F_{\mu\nu} \:\tilde{F}^{\mu\nu} = g\: a \:\boldsymbol{E} \cdot \boldsymbol{B}
    \end{aligned}
\end{equation}
where $a$ is the axion field, $ F_{\mu\nu}$ ($\tilde{F}^{\mu\nu}$) the (dual) electromagnetic field strength tensor and g the photon-axion coupling constant. This implies that in the presence of an external magnetic field photons and axions may mix \cite{7,8}. Let us recapitulate the essentials of the photon-axion mixing. For a photon traveling in the z direction, its polarization lies in the x-y plane. The component of \textbf{B} parallel to z axis does not induce any mixing. Following Equation (\ref{eq2}), the transverse magnetic field \textbf{B}$_T$ couples to \textbf{A}$_{\parallel}$, the photon  polarization parallel to \textbf{B}$_T$ and decouples from \textbf{A}$_\bot$, the photon polarization orthogonal to \textbf{B}$_T$ . The photon-axion mixing is governed by the following matrix 

\begin{equation}\label{eq3}
\begin{aligned}
M = \frac{(m_{a}^2 - m_{\gamma}^2)}{4E}N
\end{aligned}
\end{equation}

\vspace{0.2cm}

\begin{equation}\label{eq4}
\begin{aligned}
\large{N = 
\begin{bmatrix}
1 & \frac{2gBE}{(m_{a}^2 - m_{\gamma}^2)}  \\
\frac{2gBE}{(m_{a}^2 - m_{\gamma}^2)} & -1

\end{bmatrix}
}
\end{aligned}
\end{equation}
with m$_a$(m$_\gamma$) the axion mass(effective photon mass). Defining

\begin{equation}\label{eq5}
    \begin{aligned}
    tan2\theta = \frac{2gBE}{(m_{a}^2 - m_{\gamma}^2)}
    \end{aligned}
\end{equation}
we find that the eigenvalues of N are 

\begin{equation}\label{eq6}
\begin{aligned}
\lambda = \pm \: \frac{1}{cos2\theta}
\end{aligned}
\end{equation}
and the corresponding eigenstates 

\begin{equation}\label{eq7}
\begin{aligned}
V_1 = \begin{pmatrix}
cos\theta \\
sin\theta
\end{pmatrix}
\hspace{1cm}
V_2 = \begin{pmatrix}
sin\theta \\
-cos\theta
\end{pmatrix}
\end{aligned}
\end{equation}
The photon and the axion states are expressed in terms of the eigenstates. After a travel of distance z within the magnet the photon amplitude is modified to 

\begin{equation}\label{eq8}
    \begin{aligned}
    A_{\parallel}(z) = [cos^2\theta e^{-i\omega_{+} z} + sin^2\theta e^{-i\omega_{-} z}] A_{\parallel}(0)
    \end{aligned}
\end{equation}
with

\begin{equation}\label{eq9}
    \begin{aligned}
    \phi = \omega_+ - \omega_- = \frac{1}{2E}\:\Delta
    \end{aligned}
\end{equation}

\begin{equation}\label{eq10}
    \begin{aligned}
    \Delta = \frac{1}{2E}[(m_{a}^2 - m_{\gamma}^2) + 4\:g^2\:B^2\:E^2]^{1/2}
    \end{aligned}
\end{equation}

Most experiments are looking for sources of axions(solar axions, halo axions). In that case, helioscopes or haloscopes are trying to check the amplitude

\begin{equation}\label{eq11}
    \begin{aligned}
    A(a \rightarrow A_{\parallel}) = cos\theta \: sin\theta (e^{-i\omega_{+} z } - e^{-i\omega_{-} z })
    \end{aligned}
\end{equation}

In our proposal we study the coupling of the axion to a pair of entangled photons. Quantum entanglement is a distinct feature of Quantum Mechanics, supporting its non-local character \cite{9,10,11}. The wave function describing the entangled photons, named photon 1 and photon 2, is

\begin{equation}\label{eq12}
    \begin{aligned}
        \ket{\Psi} = \frac{1}{\sqrt{2}}[\: \ket{x_1, x_2} + \ket{y_1, y_2} \:]
    \end{aligned}
\end{equation}

Clearly the wave function is not factorized into a product of the individual particles. For a measurement of photon's polarization along a specific axis the result is $+1$ $(-1)$ if the polarization is found parallel to the the axis (or perpendicular to the axis). We decide to measure the polarization of photon 1 along the $x_1$ axis and the polarization of photon 2 along an axis in the $x_2 - y_2$ plane forming an angle $\beta$ with the $x_2$ axis. Quantum Mechanics offers the following results for the probabilities of joint polarizations

\begin{subequations}\label{eq13}
\begin{equation}
    \begin{aligned}
        P_{++} = \frac{1}{2} \: cos^2\beta
    \end{aligned}
\end{equation}
\begin{equation}
    \begin{aligned}
        P_{+-} = \frac{1}{2} \: sin^2\beta
    \end{aligned}
\end{equation}
\begin{equation}
    \begin{aligned}
        P_{-+} = \frac{1}{2} \: sin^2\beta
    \end{aligned}
\end{equation}
\begin{equation}
    \begin{aligned}
        P_{--} = \frac{1}{2} \: cos^2\beta
    \end{aligned}
\end{equation}
\end{subequations}

Next we introduce the magnetic field B in the $x_1$ direction. The interaction (eq. \ref{eq2}) will change the photon amplitude according to eq. (\ref{eq8}). The $\ket{x_1}$ state is replaced by

\begin{equation}\label{eq14}
    \begin{aligned}
        \ket{x_1}_a = [cos^2 \theta  e^{-i\omega_{+} z} + sin^2 \theta  e^{-i\omega_{-} z}] \ket{x_1}
    \end{aligned}
\end{equation}
The same quantum mechanical rules offers now the modified probabilities

\begin{subequations}\label{eq15}
\begin{equation}
    \begin{aligned}
        P^{a}_{++} = \frac{1}{2} \: cos^2\beta [1 - (sin2\theta \: sin \frac{\phi z}{2} )^2]
    \end{aligned}
\end{equation}
\begin{equation}
    \begin{aligned}
        P^{a}_{+-} = \frac{1}{2} \: sin^2\beta [1 - (sin2\theta \: sin \frac{\phi z}{2} )^2]
    \end{aligned}
\end{equation}
\begin{equation}
    \begin{aligned}
        P^{a}_{-+} = \frac{1}{2} \: sin^2\beta
    \end{aligned}
\end{equation}
\begin{equation}
    \begin{aligned}
        P^{a}_{--} = \frac{1}{2} \: cos^2\beta
    \end{aligned}
\end{equation}
\end{subequations}

Notice that the probabilities $P_{-+}^{a}$ and $P_{--}^{a}$ remain unchanged since they involve a photon with polarization vertical to the magnetic field. 

The ratios 

\begin{equation}\label{eq16}
    \begin{aligned}
        \frac{P^{a}_{++}}{P^{a}_{--}} = \frac{P^{a}_{+-}}{P^{a}_{-+}} = [1 - (sin2\theta \: sin\frac{\phi z }{2})^2 ]
    \end{aligned}
\end{equation}
are sensitive only to the axion parameters and the experimental setup. Thus the correlations among the polarizations of entangled photons may reveal the existence of axions. 

There are a number of advantages in our approach. In the experiments involving helioscopes or haloscopes the final outcome is dependent on models of the solar dynamics, or hypotheses on axion dark matter distribution. Our proposal is based solely on the photon-axion interaction, eq. (\ref{eq2}). There are also "light shining through a wall" experiments, based on the interaction suggested by eq. (\ref{eq2}). However one has to pay the price for a tiny conversion, the probability $P(\gamma \rightarrow a)P(a \rightarrow \gamma)$. We should notice also that the advances in laser light technology paves the way for an increased sensitivity of the proposed experiment. The experiment with the entangled photons may take place within a lab. We may consider also the propagation of satellite - based entangled photons to ground stations \cite{12}. 

Quantum entanglement lies at the heart of quantum information, quantum communication and quantum computation\cite{13}. In another direction quantum entanglement on a cosmological scale creates a geometry, which explains important aspects of our universe \cite{14}. In the present study we suggest to use quantum entanglement in order to reveal features of particle and astroparticle physics.

\end{document}